\documentclass[10pt,a4paper]{article}

\usepackage{color}
\usepackage{times}
\usepackage{graphicx}
\usepackage{comment}
\usepackage[intlimits]{amsmath}
\usepackage{ulem}
\usepackage{bm}
\usepackage{xcolor}
\usepackage{epstopdf, epsfig}
\usepackage{amsmath}
\usepackage{amsthm}
\usepackage{textcomp}
\usepackage{amsfonts}
\usepackage{amssymb}
\usepackage{float}
\usepackage{soul}
\newcommand{\HO}{H$_2$O$_2$}

\title{Metamachines of Pluripotent Colloids}

\author
{Antoine Aubret,$^{1,2\ast}$ Quentin Martinet,$^{1,3}$ Jeremie Palacci$^{1, 3\ast}$\\
\\
\normalsize{$^{1}$Department of Physics, University of California San Diego}\\
\normalsize{$^{2}$Current Address: Univ. Bordeaux, CNRS, LOMA, UMR 5798, F-33405 Talence, France}\\
\textcolor{black}{\normalsize{$^{3}$Current Address: IST Austria, Am Campus 1, 3400 Klosterneuburg, Autria}}\\
}

\date{}

\begin{document} 

\baselineskip24pt

\maketitle

\begin{abstract} 
\textcolor{black}{Machines enabled the Industrial Revolution and are central to modern technological progress: A machine's parts transmit forces, motion, and energy to one another in a predetermined manner. Today's engineering frontier, building artificial micromachines that emulate the biological machinery of living organisms, requires faithful assembly and energy consumption at the microscale. Here, we demonstrate the programmable assembly of active particles into autonomous metamachines using optical templates. Metamachines, or machines made of machines, are stable, mobile and autonomous architectures, whose dynamics stem from the geometry. We use  the interplay between anisotropic force generation of the active colloids with the control of their orientation by local geometry. This allows autonomous reprogramming of active particles of the metamachines  to achieve multiple functions. It permits the modular assembly of metamachines by fusion, reconfiguration of metamachines and, we anticipate, a shift in focus of self-assembly towards active matter and reprogrammable materials.}
\end{abstract}

The potential of mobile micromachines to probe and manipulate matter at small scales makes them a sought-after goal and a challenge for applied and basic science\cite{Palagi:2018hq}. \textcolor{black}{Their engineering encounters the challenges of assembly and reconfigurability at small scale requiring original design departing from macroscopic robots\cite{Li:2019fk,Rubenstein:2014dt}. Progress in colloidal design allows one to control assembly and shows potential to engineer dynamical micromachines. Resulting structures to date\cite{He:2020kn,Gong:2017er, Wang:2012gd, Damasceno:2012gi, Chen:2011be} are however primarily static, built from passive colloids tailored to an architecture.\newline
In contrast, energy-consuming colloids have the potential to assemble mobile structures by leveraging unusual non-reciprocal interactions arising away from equilibrium\cite{PhysRevLett.112.068301, Schmidt:2019eb, PhysRevX.5.011035, Fruchart:2021tm }. In practice, functional micromachines prominently uses concerted behavior of internally-driven parts\cite{Needleman:2017fq,Keber:2014fh, Zhang:2019eia, Maggi:2015dx, DiLeonardo:2010kka,Sokolov:2010kt,Aubret:2018cca} and monolithic templates to position active agents\cite{howse-jones-ryan-etal:2007, Bricard:2014jq,Bazant:2004bo, Ren:2019jg,Melde:2016fr} and encode function:  microswimmers power micrometric gears\cite{Maggi:2015dx, DiLeonardo:2010kka,Sokolov:2010kt}, or field-induced interactions\cite{Dreyfus:2005hc, Alapan:2019hb}. Rigid templates however come with limited reprogrammability.} 

 Light templates present spatio-temporal control with sub-micron resolution that can overcome this limitation as recently highlighted by the painting of light-powered strains of motile {\it E. Coli}\cite{Frangipane:2018cn,Arlt:2018cz}. \textcolor{black}{We demonstrated previously that light-activated and phototactic microswimmers  could self-assemble onto self-spinning microgears\cite{Aubret:2018cca}. It, however, required light-activated microswimmers that exhibit migration in gradients of light intensity, and lacked the algorithm control that can be achieved exerting optical forces on the particles.}  While optical tweezers are the workhorse of particles' manipulation at the microscale\cite{Grier-Nature-2003},  the tedious pick-and-place of all components, and permanent binding requirements plagued their use for the assembly of structures. 
 
 Here, we show the design and assembly of reconfigurable  {\it metamachines} from active colloids \textcolor{black}{using \textcolor{black}{patterns of holographic optical traps as an algorithm tool} to initiate and program geometry}. The architectures are autonomous, mobile and stable. It is made possible by harnessing the non-equilibrium properties of active particles to maintain the stability of templated architecture and achieve specified dynamics. We take advantage of the chemical and physical anisotropy of  colloidal heterodimers [Fig.1a], leading to their self-propulsion and orientation in an optical field, to  control anisotropic and reprogrammable interactions between particles. This approach overcomes traditional limitations of optical tweezers for their use in self-assembly and engineering of micromachines. 

\section*{Results}
Patterns of optical traps are placed in close-packed arrangements occupied by active heterodimers to form metamachines, that move autonomously  [Fig.1b].  The heterodimers composing the structure are physically identical but differ in their orientations as a result of their number of nearest neighbors, producing different effective interactions and forces and displaying different functions. Particles in the core sit vertically and exert attractive forces maintaining the cohesion of the structure; particles at the boundary lay flat on the substrate exerting forces that prescribe the dynamics of the structure. Following reorganization in the structure, particles reorient reversibly. It follows that particles reprogram autonomously making possible the reconfiguration and assembly of machines by merging  sub-structures. They reorganize internally to exhibit specified dynamics dictated by changes of templates [Supplementary Movie 1]. We demonstrate the design of over 40 metamachines [Fig.1c-f,  Supplementary Movies 3-5, Supplementary Fig.1], whose dynamics we rationalize. 

\paragraph{Activity-assisted Optical Trapping.} 
We \textcolor{black}{use colloidal heterodimers [Fig.1a] based on synthesis previously reported\cite{Youssef_Sacanna-NatureComm-2016} with chemical modifications improving the stability of the heterobonds to prolonged light exposure [see Methods]. They are made of a polymeric bead of 3-(Trimethoxysilyl)propyl methacrylate (TPM, radius $R_{T} \approx 0.6$ $\mu$m)  bound to a  hematite particle ($\alpha$-Fe$_2$O$_3$, radius $R_{H} \approx 0.3$ $\mu$m).} Dieletric TPM has a real index of refraction in the visible range, $n_{TPM}=1.51$,  and is chemically inert, while hematite is both absorbing and photocatalytic. Experiments are performed on a custom-made microscope, allowing superimposition of spatially-uniform blue light controlling the photocatalytic activation of hematite\cite{Palacci:2013eu} ($\lambda=425-500nm$), with patterns of holographic traps\cite{Curtis_Grier-OpticsCommunications-2002} controlled by a spatial light modulator (SLM) and a red laser at $\lambda_r = 639$ nm (waist $w_0=420nm$ and incident power $P=0.1-1$ mW), a wavelength at which hematite has high extinction cross-section and negligible photocatalytic activity [see Methods]. The heterodimers  transduce energy from chemical gradients into motion\cite{Anderson-AnnualReviewFM-1989} in aqueous solutions of hydrogen peroxide, \HO$\sim6\%$wt. \textcolor{black}{They were previously shown to propel by photocatalytic activation of hematite by blue light ($\lambda=425-500nm$) \cite{Palacci:2013eu} and exhibit phototaxis in light intensity gradients, directing themselves towards the low intensity of light\cite{Aubret:2018dq, Aubret:2018cca}.} \textcolor{black}{It follows that the wavelengths of the light used for activation (blue light) and trapping (red light) are different and that their effects can be considered independently. In the present work, we shine {\it spatially uniform} blue light, leading to isotropic propulsion of the active heterodimers with constant velocity  $v_s$ in the horizontal plane.  In effect, active heterodimers behave as generic self-propelled particles: our results and conclusions do not require light-activation and are valid for a broad class of  self-propelled particles, as further discussed below.}  

A dilute suspension of those particles  is introduced in a sealed glass capillary and sediments to a surface fraction $\sim 0.1\%$. 
Following activation, they propel along the substrate with the polymer part leading, and perform a persistent random walk\cite{howse-jones-ryan-etal:2007,Aubret:2018dq} with speed $v_s\approx 15$ $\mu$m/s and persistence time $\tau_r \approx 1$s. It leads to effective diffusion for times, $t\gg \tau_r$ with  diffusivity $D_{eff}=v_s^2\tau_r/2\sim 110\mu m^2/s$ [see Methods and Supplementary Fig.2]. 

The dynamic is drastically modified when an active heterodimer crosses an optical trap \textcolor{black}{as a result of the optical forces acting on the particle}: the particle swiftly reorients and aligns along the optical axis of the incident red light,  the polymer spheres lying on the surface as the hematite sits on top, making possible to immobilize particles according to a pattern [Fig.2a and Supplementary Movie 2].   In equilibrium, {\it i.e.} absence of photocatalytic activity or \HO\ fuel, strong scattering forces on the hematite lead to the expulsion of the heterodimer from any region of high red intensity. We measure the  lifetime $\tau$ of an heterodimer  in an optical trap at fixed propulsion speed $v_s$ and observe its increase with increased trap strength [Fig.2b] with stable trapping, $\tau>100$ s for an incident power $P=0.5$ mW, or a trap stiffness $\kappa \approx 0.8$ pN/$\mu$m obtained by conventional calibration with TPM beads [Supplementary Information].

Remarkably, the restoring force from the optical trap, $\sim \kappa w_0/2$, compares with the effective propulsive force $\gamma v_{s} \approx 0.1-0.2 $ pN of the heterodimer, $\gamma$ being the Stokes' drag, thus indicative of a balance between optical forces and self-propulsion. The reorientation of the heterodimer in a trap results from the momentum transfer between the incoming red light and the particle, set by the geometry and optical properties of the heterodimer as well as structure of the incident beam. Optical forces on the dielectric TPM component are dominated by gradient forces, as in conventional optical trapping, with a restoring force towards the center of the trap. Hematite, however, primarily experiences upward scattering forces in addition to gradient forces, due to its complex refractive index\cite{Sukhov_Dogariu-ReportsProgressPhysics-2017}. In effect, it leads to out-of-focus repulsion of the hematite and reorientation of the heterodimer by a torque $M_{scat}(\theta)$ [Fig.2c]. Following reorientation, particles propel against the radiation pressure, and stable trapping originates in propulsive forces larger than optical repulsion, leading to particles pushing against the substrate.

 To quantitatively describe the phenomenon, we perform analytic calculations\cite{RanhaNeves:19} and 3D simulations using finite elements method (Comsol). We estimate the optical forces\cite{Sukhov_Dogariu-ReportsProgressPhysics-2017} acting on the particle at different heights and distances from an incident focused beam in water, with waist $w_0 \approx 400$ nm [Fig.2d-e and Methods]. We evaluate numerically the optical torque $M_{scat}(\theta)$ for different laser power P measured in the experiment, \textcolor{black}{with increasing laser power resulting in larger aligning optical torques} [Fig.2d]. In the experiment, we further observe that heterodimers propel on the substrate with a slant,  $\theta \approx-\pi/8$ rad, as a result of phoretic and hydrodynamic interactions of the particles with their mirror-image \cite{Uspal_Tasinkevych-SoftMatter-2015,Spagniole_Lauga-JFM-2012}. The resulting torque $M_{S}(\theta)$ is  quantified from the experimental distribution of swimmers' orientation measured by simultaneous tracking of the hematite and TPM component [see Methods]. The comparison between  $M_{S}(\theta)$ and $M_{scat}(\theta)$ reveals  comparable magnitude in a range of accessible incident power [Fig.2d] and requires the inclusion of torque $M_{S}(\theta)$ to perform quantitative predictions. 
 
The interplay between particles propulsion and optical forces leads to their crossing, trapping or expulsion by an optical trap. We simulate the trajectories of non-brownian heterodimers reaching the laser spot by incorporating propulsive  and drag forces as well as $M_{scat}$ and $M_S$ torques. We summarize the predicted behavior on a phase diagram \{$P,V_{S}$\}, where stable trapping is defined by steady position of a particle within the region defined by the waist and the Rayleigh range of the incident light [Fig.2e, Methods and Supplementary Fig.7]. The quantitative agreement between the  predicted phase diagram and the experiments confirms the proposed scenario for the trapping of active heterodimers.
The trapping mechanism results from the interplay between repulsive scattering forces and propulsive forces. It is generic and applies to a broad class of active colloids, independent of the propulsion scheme: self-propelled particles do not need to be light-activated, and can be made from a  broade range of  materials. It requires that optical forces align the particles along the direction of the trap and that propulsion counteracts optical forces. We confirm the broad applicability of our approach by computing the conditions of trapping for dimers of variable size ratios, showing that larger dielectric components facilitate the trapping and suggesting experimental strategies for alternate systems [see SI and Supplementary Fig.7]. 

We further investigate the effect of shape on the scattering component and observe that in the considered optical regime, the scattering force only weakly depends on the shape of the scattering component of the active colloid and is well-approximated by the isotropic scattering by a sphere [SI and Supplementary Fig.6]. We finally compute the optical forces on a spherical Janus with a metallic cap, and confirm the existence of a restoring optical torque due to optical forces [SI, Supplementary Fig.10, 11], enabling algorithmic design of metamachines with such particles.  The accurate prediction of this Goldilocks zone of trapping for each given particle requires the quantitative estimation of the optical forces and needs to be confronted to experiments but is beyond the scope of this work.  In effect, for an active colloid with propulsion velocity $V$, it is accessible experimentally by varying the power of the incident trapping light, until optical forces balance the fixed propulsive force.

Remarkably, the enhanced diffusion of the self-propelled particles, $D_{eff} \sim 110$ $\mu$m$^{2}$, 200-fold the diffusivity of identical microparticles in equilibrium, guarantees the rapid occupancy of {\it static} traps even using dilute suspensions of active particles. It takes typically $\mathcal{A}/4D_{eff}\sim 24$s for  self-propelled particles to occupy optical traps distributed over an area $\mathcal{A} \sim 10^4 \mu m^2$  [Fig.2a, Supplementary Movie 2], thus overcoming  traditional limitations of optical manipulation of micrometric objects, thermally too slow to explore the energy landscape set by the optical traps. The non-equilibrium nature of the active heterodimers is two-fold: it allows the rapid positioning of self-propelled particles in static optical traps and effective interactions resulting from the coupling between particles' orientation and propulsive forces. It opens opportunities for assembly that we explore below.  

\paragraph{Assembly of Dynamical and Reconfigurable Metamachines.} 
As the optical traps are removed, particles initially captured in distant traps  self-propel randomly and the template vanishes. In contrast, the initial structure remains stable for particles organized in close-packed arrangements and exceeding two neighbors. Particles, which all stood vertically in the traps,  evolve differentially depending on local arrangement after removal of the optical traps [Fig.3a]. Heterodimers surrounded by 5 or 6 nearest neighbors remain vertical,  as a result of neighbors' excluded volume interactions. Facing the substrate, they generate an attractive flow\cite{Weinert:2008cn,DiLeonardo:ul}, providing cohesion to the machine. Heterodimers with 3 or 4 neighbors evolve to lay flat on the substrate [Fig.3a]. They exert forces on the structure directed along their axis of symmetry and direction of propulsion. Though the particle' orientation fluctuates in time [Fig.3b], their average orientation is primarily defined by the number of nearest neighbors as well as diffusiophoretic repulsion between clouds of chemicals from  \HO\ decomposition by hematite\cite{Aubret:2018cca,Aubret:2018dq}. Particles with 3-neighbors orient with $\bar\phi_3 \sim \pi/4$ and particles with 4-neighbors at  $\bar\phi_4 \sim \pi/2$, where $\bar\phi_n$ is the average angle made  by a swimmer with $n$ neighbors with the local tangent [Fig.3a-b and SI]. The complex distribution of orientation fluctuations [Fig.3b], as well as additional experiments using rod-like structures of various length, for which the number of nearest neighbors is conserved but the velocity of the machine is changed [Supplementary Fig.12], highlight an additional and important role of hydrodynamics. It is, however, mostly neglected to devise a simple model that predicts the dynamics of a machine from its geometry. \newline
The basic set of rules follows: particles with 1 or 2 neighbors are unstable and detach, particles with 3 or 4 neighbors exert forces on the perimeter and along their  propulsion orientation, and particles with 5 or 6 neighbors do not contribute to the motion of the architecture [Fig.3a].  It underlies a simple model that only considers particles with 3 or 4 neighbors to predict the micromachines' dynamics  from the template geometry. Each of those particles exert a point-force $F_{p,i}=\gamma V_{S}$, at the position $M_i$ of particle $i$ and along the direction $\mathbf{n_i}$ with angle $\bar{\Phi}_i$ defined primarily by the number of nearest neighbors. In addition, we assume that particles with 3-neighbors, for which the value of the angle is degenerate by symmetry, have the same angle that their nearest active 3-neighbors particle, emulating the hydrodynamic alignment observed in the experiment. We consider the machines to be solid bodies, as they do not deform when they displace. It follows that their dynamics are simply characterized by the translational velocity $\mathbf{v_G}$ of their geometric barycenter G and their instantaneous angular rotation $\Omega_{G}$, which are obtained by balancing propulsive forces with the viscous drag. Recounting that vertical particles do not contribute to the 2D dynamics, we get:
\begin{align}
\mathbf{v_G} & = \frac{V_{S}} {{N}} \sum_{ \varepsilon,i} \mathbf{n}_i\\
\mathbf{\Omega_{G}} & = V_{S}\frac{\sum_{\varepsilon,i} \mathbf{GM_i \wedge \mathbf{n_i}}}{\sum_{all,j} \mathbf{GM_j}^2}
\label{eq:1}
\end{align}
where $\sum_{{ \varepsilon}}$ is the sum over the active edge, {\it i.e} swimmers with 3 or 4 neighbors and $\sum_{all}$, the sum over all  $N$ particles constituting the structure. Our simple model qualitatively captures the measured dynamics for $22$ different colloidal architectures but overestimates rotation speeds by discounting fluctuations of orientation [Fig.3c-inset]. To account for their role, we record simultaneously the dynamics of a structure as well as the instantaneous positions and orientations of all constitutive particles. We observe quantitative agreement between the measured and predicted rotational [Fig.3c] and translational speed [Fig.3d] for all time and for all structures, confirming the validity of the model and the role of geometry in programming the dynamics of a metamachine. 

Remarkably, interactions between particles do not result from intrinsic properties but rather from the interplay between geometric arrangement and self-propulsion. As they reorient reversibly, otherwise identical particles reprogram autonomously to different functions in response to changes in the structure or their environment [Fig.4a]. The ability for particles to reprogram autonomously allows reconfiguration and provides unmatched control to dynamical assembly. It  highlights a salient advantage of the proposed strategy and contrasts with conventional  assembly encoded in properties of the colloidal building blocks, e.g. shape\cite{Sacanna:2011dd} or specific interactions\cite{He:2020kn, Wang:2012gd,Chen:2011be}. This feature is harnessed to assemble and reconfigure bespoke machines from a set of identical constitutive building blocks.  For example, we devise a slender structure by template-and-merge of substructures, requiring particles to reprogram autonomously: particles previously sitting at the edge of templated metamachines become integrated to the interior of a slender structure [Fig.4b]. They spontaneously reorient  to stand vertically, simultaneously annealing defects between merged substructures and providing cohesion to a novel architecture. It autonomously rotates. The architecture is reconfigured by application of a novel template: particles from the edge are ejected and a novel active boundary forms. Particles reprogram so that  the particles now on the boundary reorient and contribute to the rotation of the structure [Fig.4b, Supplementary Movie 1]. 
 
We further design and assemble a broad range of colloidal metamachines. Chiral structures that rotate predictively [Fig.4c,d], or rotationally symmetric structures, which direction of rotation results from spontaneous symmetry breaking\cite{Aubret:2018cca} following removal of the template [Fig.1c, Supplementary Movie 3]. In effect, only convex structures are accessible from direct templating:   wrongful occupation of vacant positions in concave shapes due to attractive interactions otherwise prescribe the formation of the desired structures. We overcome this limitation using the spatiotemporal control offered by the SLM. We first assemble metamachines from convex templates and combine them with individual particles to obtain a stable dynamical architecture with concave edges [Fig.4c, e]. This template-and-merge allows the formation of more complex machines: rotators from centro-symmetric objects, whose chirality results from the conversion of a particle on the edge with 3 neighbors (and exerting forces) into a particle with 5 neighbors solely contributing to the cohesion of the structure [Fig.4c, d]. The control offered by the SLM allows one to discriminate the particles undergoing activity annihilation and control the chirality of the final architecture from identical substructures. In order to demonstrate the broad applicability of the strategy, we  devise  a translating metamachine \textcolor{black}{from the fusion and autonomous reconfiguration of rotating structures. It further presents} a physical cavity to demonstrate the opportunities of metamachines to manipulate objects at the microscale [Fig.4e-f and Fig.1e, Supplementary Movie 4]. As effective interactions exert generically on all particles, we can program hybrid metamachines, using also optical patterns and passive particles as templates. The active heterodimers provide provide cohesion and force generation to the hybrid metamachine [Fig.1f and Supplementary Movie 5]. This further highlights the potential of pluripotent active colloids for self-assembly and provides a natural point of convergence between our design of metamachines and state-of-the-art colloidal design. 

\section*{Discussion}
This work offers a unique outlook for self-assembly. It harnesses forces exerted by active building to devise multi-fated particles, with reprogrammable (effective) interactions, that evolve reversibly and autonomously. \textcolor{black}{It leverages optical traps as an algorithmic tool to program architecture and reinstates optical manipulation as a darling of assembly at the microscale by overcoming its equilibrium limitations. The algorithm is generic and will be applicable to a broad variety of active particles, and notably Janus microswimmers [see SI for discussion]. The study of the autonomous response and reconfiguration of the architectures under environmental changes will open opportunities of complex time-dependent dynamics, that go beyond the scope of this work}. Our approach will further benefit from progress in colloidal design\cite{He:2020kn,Gong:2017er} to advance self-assembly. It also provides uncharted opportunities for the development of shape-shifting materials\cite{Nguyen:2011jg, Banerjee:2017dc} and the development of machines made of machines.


\newpage
\section*{Methods}

\subsection{Synthesis of hematite cubes}
The synthesis procedure of hematite cubes follows the sol-gel method described by Sugimoto\cite{Sugimoto_Mochida-JCIS-1998}. Briefly, we mix 100 mL of 2M FeCl$_3$ $\cdot$ 6H$_2$O, 90 ml 6 NaOH and 10 ml water, in a 250 mL pyrex bottle and shake thoroughly. Immediately after, the bottle is placed in an oven at $100^\circ$C, and aged for 3 to 4 days, until the hematite particles reach the desired size (growth is monitored by optical microscopy). The resulting hematite cubes in the gel network are isolated by successive sedimentation and resuspension cycles in DI water [Supplementary Fig.2].

\subsection{Synthesis of heterodimers}
The synthesis of heterodimer particles is performed by heterogeneous nucleation of trialkoxysilanes (oil precursors) on solid hematite particles as seeds, and the control of their wetting properties with the pH of the solution. The synthesis procedure is adapted from ref.\cite{Youssef_Sacanna-NatureComm-2016}, with chemical modification to reinforce the stability of the heterodimer under light illumination. In particular, we make use of a hydrophobic copolymer Hexadecyltrimethoxysilane (HTS) to chemically protect the bond between the hematite and polymer core against highly reactive hydroxil radicals generated during H$_2$O$_2$ consumption. A beaker with 100 mL of DI water is prepared, and mixed with 120 $\mu$L of a 50\% NH$_3$ solution, giving a pH $\sim 10.5$. The solution is kept under mild magnetic stirring. We add $\sim 1$ mL of an aqueous solution of hematite particles, to get a slightly red-colored solution. Following, we add 100 $\mu$L of HTS, immediatly followed by 500 $\mu$L of 3-(Trimethoxysilyl)propyl methacrylate (TPM). The solution is then covered with parafilm, and let under mild stirring for $\sim$ 2h00. During this time, the HTS and TPM hydrolyse, and condense on the hematite particles, with strong wetting leading to their engulfment in the oil phase. After $\sim$ 2h00, the solution is turbid, as a result of the scattering of newly formed colloidal particles. We then add 2 mL of Pluronic F-108 solution (5\% wt), and wait 2 mn. The dewetting and extrusion of TPM from the hematite is performed by changing the pH of the solution to pH $\sim$ 2.1, by adding 1.5 mL of 1M chloridric acid HCl. The solution is let under stirring for 3 mn, and diluted 4 times. We then carry out the polymerization by adding 50 mg of radical initiator Azobisisobutyronitrile (AIBN) to the solution and leave it under stirring for $\sim 5$ mn. The beaker is covered with an aluminium foil and placed in a pre-heated oven at $\sim$ 80 degrees Celsius for 2 hours. We let the solution cool down to room temperature, remove the excess solution above the sedimented particles, and add 50 mL DI water with 1 mL of 250 mM NaOH solution, giving a pH $\sim$ 10, and let the solution overnight to facilitate the hydrolysis and condensation of any remaining HTS monomers. The solution is then centrifuged and rinsed multiple times to remove the excess TPM/HTS particles and obtain the desired colloidal solution of heterodimers. 

 \subsection{Sample preparation}
 The samples are prepared at low particle density of $\sim 10^{-3}$ $\mu$m$^{-2}$. The heterodimer particles are diluted in a 6\% solution of hydrogen peroxide H$_2$O$_2$ (Fisher Scientific H325-500) in deionized water (Milli-Q, resistivity 18.2M). The cell for the solution is composed of a rectangular capillary glass (VitroCom 3520-050), previously subjected to plasma cleaning (Harrick Plasma PDC-001) and thoroughly rinsed with DI water. The solution is injected in the capillary, then sealed with capillary wax (Hampton Research HR4-328). As the particles are non-buoyant, they sediment near the bottom surface of the capillary, and observation with the microscope is made in this plane.

\subsection{Optical setup}
 The experiments are carried out on a custom-made optical setup, allowing for simultaneous uniform excitation of the microswimmers and holographic optical tweezing, as presented on [Supplementary Fig.3]. The sample is observed with bright-field transmitted illumination (LED1). A LED is set up in the blue range (LED2, $\lambda = 425 - 500
$ nm, Lumencor SOLA 6-LCR-SC) and uniformly illuminates the sample on a large area (size $\sim 300 \times 300$ $\mu$m$^2$) to activate the swimmers by triggering the photocatalytic decomposition of H$_2$O$_2$ by the hematite (typical intensity of $\sim 1 \, \mu$W/$\mu$m$^2$). A red continuous laser with near TEM00 mode ($\lambda = 639$ nm, Coherent, Genesis MX639-1000 STM, $M^2 < 1.1$) is added on the optical path. The linearly polarized beam is collimated, rotated with a $\lambda/2$ waveplate, and sent on the surface of a Spatial Light Modulator (SLM, Holoeye -LETO Phase Only Spatial Light Modulator). The optical path follows a typical 4-f setup using two $f_1=400$ mm lenses, and the zero-order of the diffracted beam is filtered out with a diaphragm at equal distance $d=f$ between the two lenses. Following, a hologram is formed at the back aperture of a high NA objective (Nikon apo-TIRF, $\times 100$, NA=1.45) allowing for the creation of complex spatiotemporal optical patterns in the object plane, at the bottom surface of the sample. The hologram is computed in real time using a computer software (Holoeye), with the phase patterns computed under Matlab, and allows for the selective trapping and manipulation of individual swimmers. 
An electronic shutter (Thorlabs SHB1T) on the red optical path enables switching ON and OFF the laser traps.
 The sample is mounted on a manual micrometric stage (Nikon Ti-SR). Observation is performed through the same objective as for excitation, and the bright-field signal is reflected on a polarizing beam splitter and observed on 2 monochrome Charged Coupled Devices (CCDs) with different resolutions (0.05 $\mu$m/px and 0.1 $\mu$m/px, respectively, Edmund Optics EO-1312M), with appropriate spectral filters.

\subsection{Image acquisition and analysis} : All experiments were recorded on the CCD camera with 0.05 $\mu$m/px resolution (except for Supplementary Movie 2 performed at lower magnification), at frame rate between 20-50 fps, and under bright field illumination. Tracking was performed separately for the hematite and TPM particles with standard Matlab routine.
For each of the 22 investigated structures, we record at least $10$ s of experiments. The positions are then averaged a first time over $200$ ms to compute $\omega(t)$, $\Omega_G(t)$, $\mathbf{v}_G(t)$. We average again those points to obtain 4 data points with equivalent statistical weight for rotationnal motion for each individual structure [Fig.3c]. For translational motion, all data points for all structures are averaged together to obtain the desired graph. Error bars are obtained from standard deviations of experimental measurements.

\subsection{Analytical calculation of the optical field}
The system is in the  Mie regime, and uses a strongly focused laser beam (NA $\sim 1$). We therefore use exact formulas for the optical forces acting on an a homogeneous sphere of any composition in the Mie-regime, formalized upon the Generalized Lorenz-Mie Theory (GLMT) for a highly focused incident gaussian beam in the angular spectrum representation. 
Knowing the full expressions of the scattered fields, the optical forces are obtained by integrating the Maxwell Stress tensor over a surface surrounding the particle. The full expressions of $F_x$, $F_y$, and $F_z$ and their complete derivation is described in Ref.\cite{RanhaNeves:19}. 
We consider an incident beam at $\lambda=639$ nm filling the entire back aperture of a water (refractive index $n_{w}=1.33$) immersion objective, polarized along X and propagating along the Z-direction. We chose the parameters of the optics as to get a waist of $\omega_0 \approx 400 nm$ that closely matches the experimentally determined value [Supplementary Fig.8]. The TPM part of the heterodimer is modeled as a non-absorbing dielectric bead of refractive index $n_{T} = 1.51$ and radius $R_{T}=500$ nm, and the hematite part is modeled as a spherical body of radius $R_H=250$ nm, homogeneously filled with $\alpha-$Fe$_2$O$_3$ ($n_H=3.10+0.065i$ at $\lambda=639$ nm).
 The analytical expression, from Ref.\cite{RanhaNeves:19}, calculates the forces on individual TPM and hematite particles, that we use to approximate the total force on a heterodimer [Supplementary Fig.5].

\subsection{Numerical calculation of the optical field}
		\label{sec:comsol_calc}
	We use Comsol Multiphysics (v5.5) to compute the 3D optical field generated by a single, focused laser beam in presence of a \textit{full} heterodimer particle in water. The solution for the background field is analytically approximated using the angular spectrum representation of the focal field for a non-paraxial beam, and using discrete summation to approximate the continuous integration over the numerical aperture of the focusing lens.

	We model infinite boundaries in our system using the Perfectly Matched Layers module of COMSOL, which prevents any reflection of the scattered field on our artificial boundaries.
	The optical forces $F_x$, $F_y$, $F_z$ are then obtained by integrating the Maxwell stress tensor on the surface of the particles\cite{Sukhov_Dogariu-ReportsProgressPhysics-2017}. We check the incident power by integrating the flux of the Poynting vector associated to the background field across the transverse XY plane.
	
	The results are compared to the total force obtained by summation of the forces exerted on individual TPM and hematite particles, computed using analytical expressions [Supplementary Fig.5]. We show that the total force is well approximated by individually calculating the optical forces on a single hematite and a single TPM particle, and summing them after.
	We note that for single particles, the forces computed with COMSOL agree perfectly with the theory above for identical beam parameters.	We therefore use this method for mapping of the optical forces in the $\{X,Z,\theta\}$ space.
	
\subsection{Motion of a self-propelled heterodimer entering an optical trap - simulations}
We consider a simplistic model where the heterodimers are self-propelled particles in the absence of noise. Heterodimers are considered rigid bodies constituted of 2 distinct parts in the XZ plan, with orientation $\mathbf{u_{\theta}}$.  The position $\mathbf{r}=\{x,y=0,z\}$ is defined as $\mathbf{r} =[\gamma_H \mathbf{r_H} + \gamma_T \mathbf{r_T}]/(\gamma_H + \gamma_T)$ to account for the different drag coefficients $\gamma_H=6\pi\eta R_H$ and $\gamma_T=6\pi\eta R_T$ acting on the hematite and TPM beads situated at $\mathbf{r_H}$ ad $\mathbf{r_T}$, respectively. In our model, we discard hydrodynamic interactions between the hematite and the TPM. The equations of motion are determined from balancing external forces and self-propulsion by the viscous drag, i.e. : $\sum \mathbf{F}_{ext} = \mathbf{F}_{drag} $ and $\sum \mathbf{M}_{ext} = \mathbf{M}_{drag}$.

The total drag torque is obtained as ${M}_{drag} = \Gamma \omega$, with $\Gamma = \gamma_H {R_T}^2+\gamma_T{R_H}^2 + (4/3)(\gamma_H R_H^2+\gamma_c R_T^2)$. For a swimmer with propulsion velocity $V_S$, we write :
\begin{align}
\frac{dx}{dt} & =V_S \cos(\theta) +\frac{\sum F_{ext,x}}{\gamma}\\
\frac{dz}{dt} & =V_S \sin(\theta) +\frac{\sum F_{ext,y}}{\gamma}\\
\frac{d\theta}{dt} & =\frac{1}{\Gamma} \sum M_{ext}
\label{eq:motion}
\end{align}
where $\gamma=\gamma_H+\gamma_T$.

We simulate the motion of particles in the presence of optical forces, acting independently on the particles, using finite-difference equations as derived from eq.3-5.
We consider the linear optical regime, such that the optical forces $F_{opt} \propto P_{inc}$, with $P_{inc}$ the incident power. We model the substrate as an impenetrable wall situated at $z=-R_T$, such that any step for which the particle penetrates the wall yields to a displacement of the swimmer to be in close contact with the wall, i.e. we force $z_T \geq 0$, and $z_H\geq R_H-R_T$. We obtain $M_{S}(\theta)$ by interpolation from the experimental values, and apply the same procedure for the optical potential $M_{scat}(\mathbf{r},\theta)$ : we extract the forces applied on the hematite and TPM parts individually, allowing us to approximate both torques and forces $F_{scat,x,z}(\mathbf{r},\theta)$ acting on the heterodimer.

The torque $M_S(V_S,\theta)$ depends on the instantaneous propulsive force (i.e velocity $V_S$) of the swimmer. In the Stokes regime ($Re<<1$), we can reasonably assume both phoretic and hydrodynamic effects to be linearly dependent on the velocity of the swimmer. Hence, we assume linear dependency of $M_S(V_S,\theta)$, such that $M_S(V_S,\theta)=(V_S/V_{S,0}) M_S(\theta)$, where $V_{S,0} = 16$ $\mu$m/s is the velocity of the swimmer in free space.
Finally,  the  gravitational torque is negligible to the typical values for $M_{scat}$ and $M_S$, and the effect of the gravitational torque is therefore not included when computing the trajectories.

\subsection{Data availability}
The data that support the plots within this paper and other findings of this study are available from the corresponding authors upon request.

\subsection{Code availability}
The code that support the plots within this paper and other findings of this study are available from the corresponding authors upon request.

\newpage

\bibliographystyle{naturemag}

\section*{Acknowledgments}
 The authors thank R. Jazzar for useful advice regarding the synthesis of heterodimers. We thank S. Sacanna for critical reading. This material is based upon work supported by the National Science Foundation under Grant No. DMR-1554724 and Department of Army Research under grant W911NF-20-1-0112.

\section*{Author Contributions}
A. A. and Q.M. performed the experiments. A. A. analyzed the experimental results.  A.A. and J.P. conceived the project, designed the experiment, developed the model and wrote the manuscript. All authors discussed the results and commented on the manuscript.

\section*{Competing Interests} 
The authors declare that they have no competing financial interests.

\newpage

\begin{figure}
\centering
\includegraphics[scale=.8]{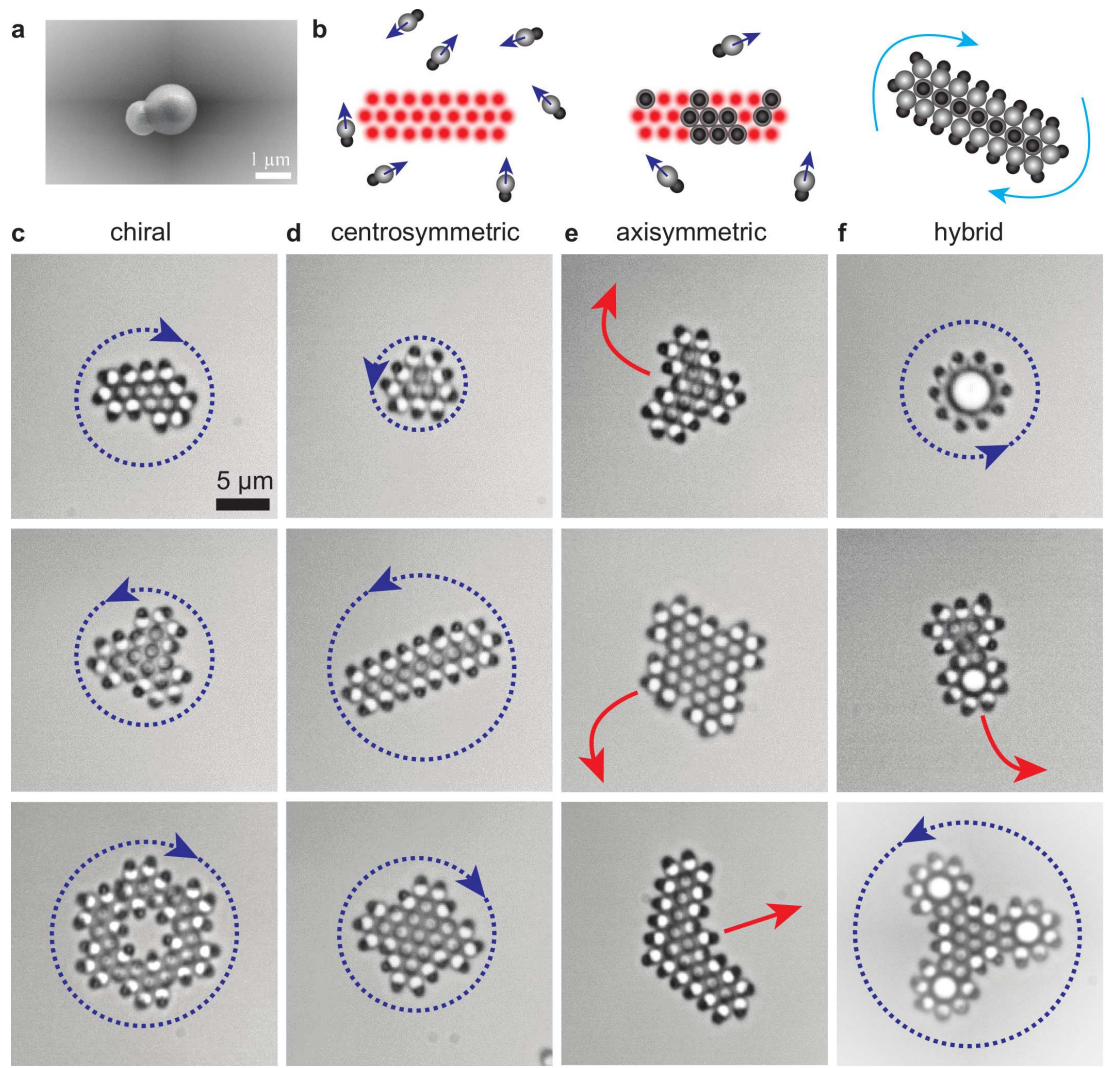}
\label{Fig1}
\caption{\footnotesize\textbf{Colloidal metamachines}. {\bf a.} Scanning Electron Microscopy of a TPM-hematite heterodimer. {\bf b.} Optical assembling of a metamachine. An optical pattern of traps is superimposed to the blue illumination activating the heterodimers. They travel randomly on the substrate until they encounter and optical trap and reorient vertically. In absence of propulsion, heterodimers are repelled by scattering by  trapping laser. Upon removal of the optical template, the structure is stable, particles reorient and the structure exhibits autonomous dynamics [see Main Text]. {\bf c-f.} Bright field pictures of colloidal metamachines. {\bf c.} Chiral metamachines rotate in fixed direction, {\bf d.} Centro-symmetric metamachines rotate by spontaneous symmetry breaking. {\bf e.} Axisymmetric metamachines with translational motion. {\bf f.} Hybrid metamachines composed of passive spheres and active heterodimers. Blue arrows indicate the direction of rotation, red arrows indicate the direction of propulsion.  Identical scale for all images.}
\end{figure}

\begin{figure}
\centering
\includegraphics[scale=.8]{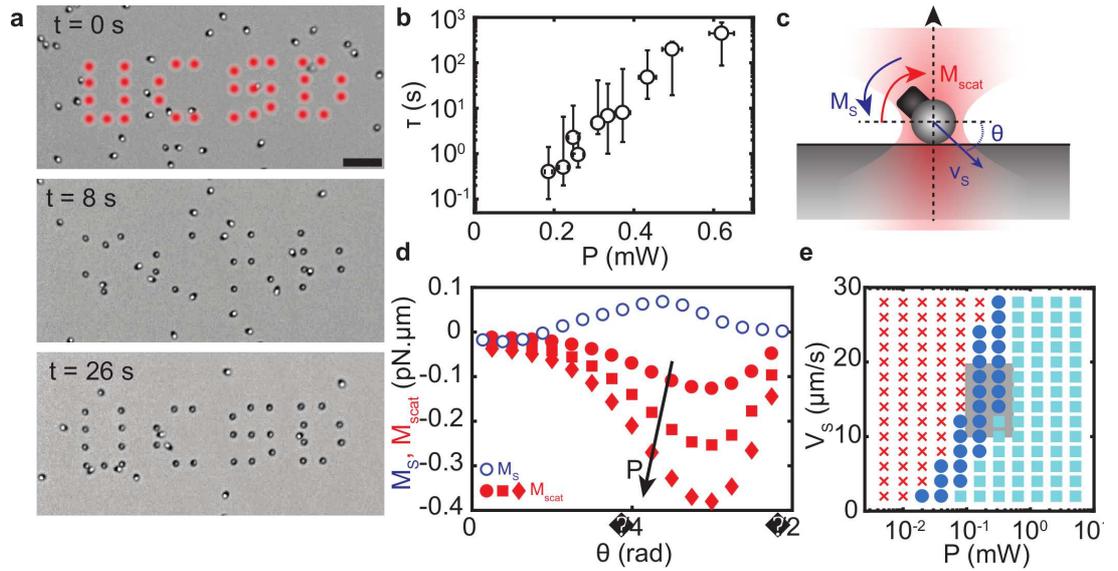}
\label{Fig2}
\caption{\footnotesize\textbf{Activity-assisted optical trapping}. {\bf a.} Time lapse of active particles in a pattern of optical traps forming the letters UCSD.  Enhanced diffusivity of self-propelled particles leads to rapid occupancy of the optical traps. Scale bar is $10$ $\mu$m. {\bf b.}  Trapping lifetime $\tau$ of a microswimmer measured at different incident laser power $P$. Stable trapping is achieved for $P \gtrsim 0.5$ mW. \textcolor{black}{Symbols are the median values and the error bars reflect the extremal values found.}
 {\bf c.} Trapping process. Gradient forces pull the particle at the center, while scattering forces on the hematite exert a torque $M_{scat} (\theta)$ on the heterodimer. It is balanced by $M_S(\theta)$ resulting from the interaction of the swimmer with the substrate. Self-propulsion at speed $V_S$ opposes scattering forces and leads to stable trapping near focus. {\bf d.} Experimental measurement  of the torque $M_S$ (empty blue circles), and numerical determination  of $M_{scat}$ (full red symbols) for a swimmer with the TPM bead at the center of the focus for increasing incident  power  $P$: 0.25 (circles), 0.5 (squares), and 0.75 (diamonds) mW. {\bf e.} Simulated phase diagram $\{P,V_S\}$ for active heterodimers entering a trap with incident power $P$ with velocity $V_S$, in the absence of noise. The outcome results from the interplay between activity and optical forces. When reaching an optical trap, a swimmer can cross it (red crosses), be ejected by radiation pressure (cyan squares), or be trapped (blue circles). The grey rectangle overlays the range of values in the experiment, showing quantitative agreement with the numerical model.}
\end{figure}

\begin{figure}
\centering
\includegraphics[scale=.8]{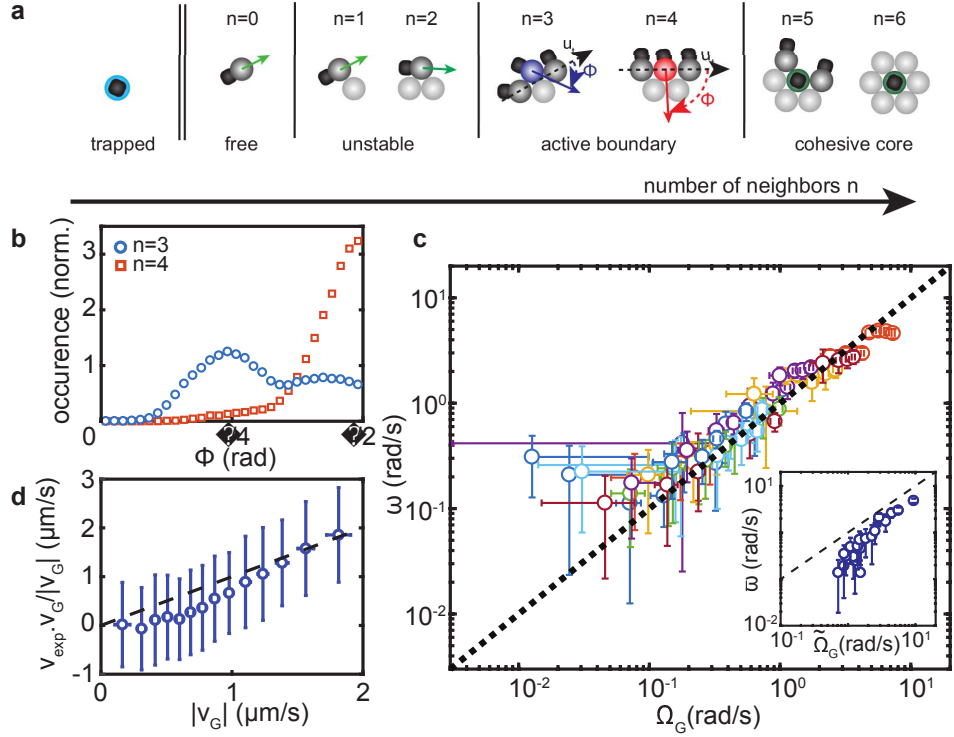}
\label{Fig3}
\caption{\footnotesize\textbf{Pluripotent colloids.} {\bf a.} Heterodimers exhibit different functions as a function of  their number of neighbors $n$. Heterodimers stand vertically in an optical trap (light blue). For $n = 0$ heterodimers navigate freely.  For $n=1$ or $2$, the structure is unstable and the heterodimers detach. For $n=3$ (purple) or $4$ (red), heterodimers orient with respect to the local tangent $\mathbf{u_t}$, with $\bar{\Phi}_3 \sim \pi/4$ and $\bar{\Phi}_4 \sim \pi/2$, respectively and exert a force along their propulsion direction.  For $n=5$ or $6$ (green), heterodimers are vertical and exert attractive interactions. {\bf b.} Histograms of the angles $\Phi$ of heterodimers for $n=3$ (blue circles) and $n=4$ (red squares) obtained from analysis of slender micromachines of different lengths. {\bf c.} Measured angular velocities $\omega$ for 22 different machines (1 color and 4 data points per machine), plotted against the prediction $\Omega_G	$ obtained by tracking of the instantaneous orientation and position of all constitutive swimmers in eq.\eqref{eq:1}. {\bf c-inset.} Mean value $\bar\omega$  plotted against  $\bar\Omega_G$ predicted from our simple model with fixed orientations $\Phi(n=3)=\pi/4$ and $\Phi(n=4)=\pi/2$ [see Main Text]. {\bf d.} Projected translational speed measured by averaging data from all 22 structures plotted against the predicted translational speed $\mathbf{v_\mathrm{G}}$  from eq.\eqref{eq:1}. Black dashed lines in C and D indicate unity slopes. \textcolor{black}{ Error bars are one standard deviation.}}
\end{figure}

\begin{figure}
\centering
\includegraphics[scale=.8]{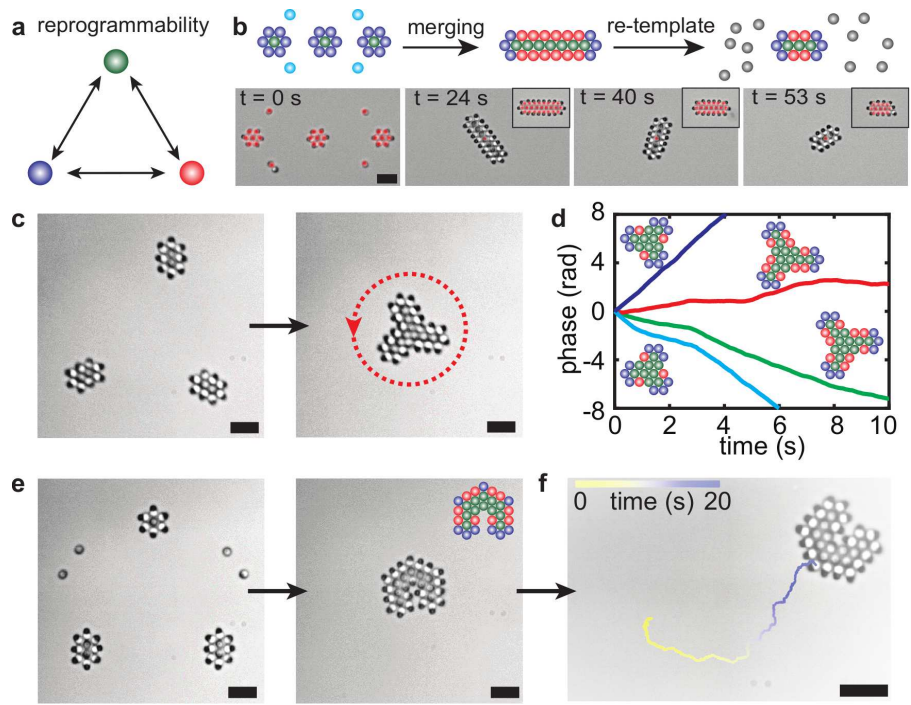}
\label{Fig4}
\caption{\footnotesize\textbf{Reprogrammable Metamachines}.  {\bf a.} Heterodimers reprogram autonomously and change functionality (colors as in Fig. 3a). {\bf b.} Metamachine Reconfiguration. \textit{(top)} : Schematic of the reconfiguration process. Colors indicate the functions as defined in Fig.3a. They highlight the autonomous reprogrammation of the particles as the structure reconfigures. Substructures are templated then merged to build a slender metamachine. Changes in templates lead to reconfiguration of the machine enabled by the autonomous reprogrammation of the constitutive particles. \textit{bottom}: Experiment time-lapse. Red circles indicate optical traps, used for re-templating and practical pinning of the structure. Insets : optical patterns for re-templating. {\bf c.}  Formation of chiral structures by template-and-merge. The dashed arrow indicates the direction of rotation.  {\bf d.} Rotational dynamics of 4 different chiral structures with diagrams highlighting the role of red ($n=4$) swimmers to set the chirality.  {\bf e.} Assembly  of a translational metamachine by template and merge. The diagram highlights the role of green ($n=5$) swimmers at the concave angle to  introduce the forward/backward asymmetry and persistent self-propulsion of the machine, presented in  {\bf f.}. Scale bars are $5$ $\mu$m.}
\end{figure}

\end{document}